\documentclass{article}

\PassOptionsToPackage{sort,square,comma,numbers,compress}{natbib}

\usepackage[preprint]{neurips_2021}

\usepackage[utf8]{inputenc} 
\usepackage[T1]{fontenc}    
\usepackage{hyperref}       
\usepackage{url}            
\usepackage{booktabs}       
\usepackage{amsfonts}       
\usepackage{nicefrac}       
\usepackage{microtype}      
\usepackage{xcolor}         
\usepackage{graphicx,wrapfig}

\usepackage[linewidth=1pt]{mdframed}
\usepackage{tikz-dependency}

\usepackage{graphicx}
\usepackage{color}
\usepackage{wrapfig}
\usepackage{listings}
\usepackage{graphbox}
\usepackage{caption}
\title{Explainability Pitfalls:\\ Beyond Dark Patterns in Explainable AI }

\author{
Upol Ehsan \& Mark O. Riedl\\
Georgia Institute of Technology, Atlanta GA, USA\\
\texttt{\{ehsanu,riedl\}@gatech.edu
\vspace{-10pt}}

}

\begin{document}

\maketitle

\begin{abstract}
To make Explainable AI (XAI) systems trustworthy, understanding harmful effects is just as important as producing well-designed explanations.
In this paper, we address an important yet unarticulated type of negative effect in XAI.
We introduce \textit{explainability pitfalls} (EPs), unanticipated negative downstream effects from AI explanations manifesting even when there is \textit{no intention} to manipulate users. 
EPs are different from, yet related to, {\em dark patterns}, which are intentionally deceptive practices.
We articulate the concept of EPs by demarcating it from dark patterns and highlighting the challenges arising from uncertainties around pitfalls.
We situate and operationalize the concept using a case study that showcases how, despite best intentions, unsuspecting negative effects such as unwarranted trust in numerical explanations can emerge.
We propose proactive and preventative strategies to address EPs at three interconnected levels: research, design, and organizational.
\end{abstract}

\section{Introduction}
Safety and trustworthiness are major goals of Explainable AI (XAI), a research area that aims to provide human-understandable justifications for the system's behavior~\cite{adadi2018peeking,ehsan2019automated,guidotti2018survey}.
This safety is vital because AI systems increasingly power decision-making in high-stakes domains like healthcare~\cite{holzinger2017we,katuwal2016machine,che2016interpretable,loftus2020artificial}, finance~\cite{MacKenzie2018, murawski2019mortgage},  and criminal justice~\cite{Rudin2020, Kleinberg2017, hao2019jail}.
Adding explainabilty does not necessarily guarantee positive effects; there can also be detrimental ones. 
To facilitate AI safety, beyond designing effective explanations, we also need to properly understand detrimental effects of AI explanations.

Despite commendable progress in XAI, emerging work has highlighted detrimental effects of explanations~\cite{stumpf2016explanations,smith2020no,Kaur2020,ghai2021explainable}.
For instance, deceptive practices can be intentionally used to design placebic explanations (devoid of justificatory content) and engender trust in AI systems and obfuscate harm~\cite{Eiband2019}. 
\textit{However, not all detrimental effects are intentional.}
Despite the designer's best intentions, it is possible to design explanations that have unintentional negative effects on the end-user.
In such cases, what can we do? How might we conceptualize intentional negative effects of explanations? When it comes to conceptualizing negative effects of explanations, especially ones that are \textit{not intentional}, there is a scarcity of conceptual artifacts.

To address this problem, we introduce the concept of \textbf{explainability pitfalls} (EPs), which are unanticipated and unintended negative downstream effects from AI explanations that can cause users to act against their own self-interests, align their decisions with a third party, or exploit their cognitive heuristics.
Examples of these downstream negative effects include user perceptions like misplaced trust, over (or under) estimating the AI's capabilities, and over-reliance on certain explanation forms. 

Explainability pitfalls (EPs) are different from \textit{dark patterns} (DPs), which are a set of deceptive practices \textit{intentionally} and “carefully crafted to trick users into doing things...They do not have the user’s interests in mind”~\cite{brignull2015dark}. 
Emerging work showcases how dark patterns can manifest in XAI to create a false sense of security and  tricking users into over-trust systems~\cite{chromik2019dark}. 
With roots in UX design~\cite{gray2018dark}, DPs have been explored in multiple contexts like games~\cite{zagal2013dark} and medical records~\cite{capurro2021dark}.
A major difference between EPs and DPs is the difference in \textit{intentionality}---DPs have  bad-faith actors intentionally trying to trick users. 
Despite being different, EPs and DPs are related; you can turn pitfalls into dark patterns by intentionally setting the traps (to trick the user).

We use the metaphor of ``pitfalls'' to signal unsuspected or hidden difficulties or dangers that are not easily recognized. 
These difficulties might arise due to lack of information, understanding, or oversight (or a combination of these). 
Without the awareness of how to identify and avoid pitfalls, there are increased risks for end-users who may never be aware of being affected.
Moreover, explainability pitfalls may only be symptomatic, thus detectable, \textit{after} users interact with the explanations and there is a misalignment between their behavior and designers' expectations.
This implies is a high bar of accountability for designers to be ``pitfall-aware'' when designing XAI systems (at first glance, the lack of intentionality in EPs may be misinterpreted as an exonerating force).
Taking the metaphor of pitfalls further, we can envision navigation strategies to detect and avoid them.
If we are pitfall-aware in our navigation of the design space and aware of the possibility of unanticipated and unintentional negative effects, we can proactively build resilience against the pitfalls.

The \textit{motivation} for conceptualizing EPs is not purely theoretical or speculative. 
It is practically motivated and empirically situated in our prior work~\cite{ehsan2021explainable} that (amongst other things) showcased how, despite no intentions to trick anyone, unsuspecting negative effects can emerge from interpretations of AI explanations. The conceptual introduction of \textit{explainability pitfalls} aims to 
(1)~bring awareness to previously unrealized intellectual blind spots (around negative effects of AI explanations), which, in turn, can
(2)~expand the XAI design space.  
This paper is not a comprehensive treatise of EPs; rather, it takes a foundational step towards operationalizing the notion both conceptually and practically.

Below we operationalize the concept of EPs by situating it through a case study where EPs manifested and were discovered through qualitative analysis of end-user responses to explanations in a controlled setting.
Reflecting on the findings, we then propose formative strategies to address EPs.


\section{Case Study: Situating Explainability Pitfalls}
\label{sec:case}
\vspace{-5pt}

In this case study~\cite{ehsan2021explainable}, we investigated how two different groups---people with and without a background in AI---perceive different types of AI explanations.
We probed for user perceptions on three types of AI-generated explanations: 
(1) natural language with justification, 
(2) natural language without justification, and 
(3) numbers that provide uncontextualized transparency into agent’s actions. 
For our purposes to situate the notion of EPs, we need to focus on how and why both groups reacted to the numbers from the Numerical-Reasoning (NR) robot (\#3).

\begin{wrapfigure}[21]{r}{0.35\textwidth}
    \centering
    \vspace{-10pt}
    \includegraphics[width = 0.345\textwidth]{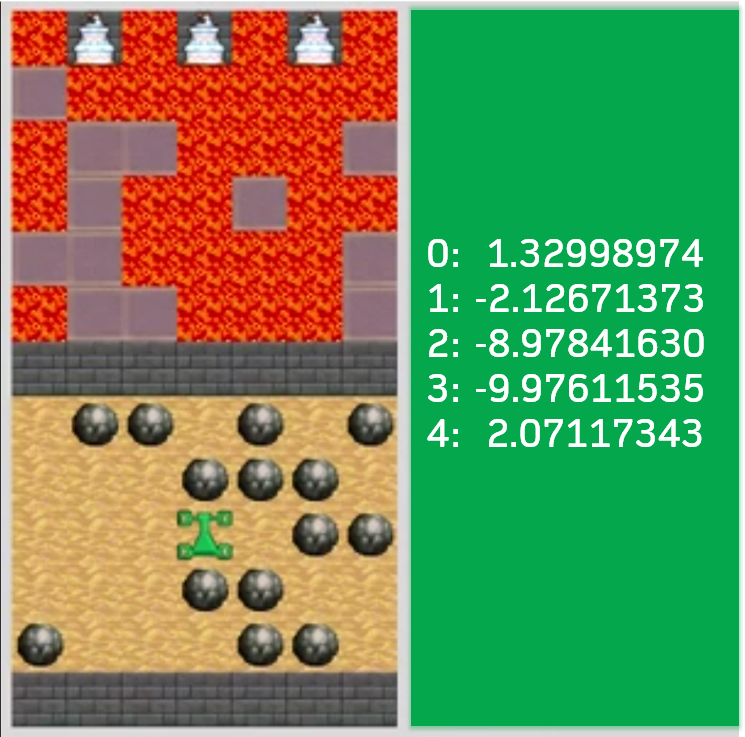}
    \caption{Screenshot depicting the Numerical-Reasoning (NR) robot navigating the task environment and ``thinking out loud'' using numbers. Taken from~\cite{ehsan2021explainable}.
    This robot was used as a foil against two other robots with natural language explanation strategies.
    \vspace{-10pt}}
    \label{fig:NRrobot}    
\end{wrapfigure}

In the study, participants provided both qualitative and quantitative perception information after watching videos of three robots (AI agents) using Reinforcement Learning to navigate a sequential decision-making environment---a field of rolling boulders and flowing lava to retrieve essential food supplies for trapped space explorers (more task details in~\cite{ehsan2021explainable}). 
The robots behave identically except in the way they “think out loud” or explain themselves. 
The NR robot (the relevant one for this paper) “thinks out loud” by simply outputting the numerical Q-values for the current state (Fig.~\ref{fig:NRrobot}).
Q-values~\cite{watkins1992q} can provide some transparency into the agent’s beliefs about each action’s relative utility (“quality”), but do not contain information on “why” one action has a higher utility than another.
Participants were \textit{not} told that the numbers are Q-values nor do they know which values correspond to which actions.

We found that \textit{participants in both groups had unwarranted faith in numbers}. However, their extent and reasons for doing so were different. 
To understand the reasons behind the effect, we will use the notion of cognitive heuristics (rules-of-thumb or mental short-cuts), which leads to biases and errors if applied inappropriately~\cite{kahneman2011thinking,wason1974dual,petty1986elaboration}. 
We will see how different heuristics, tied to one’s AI background, can lead to undesired outcomes. 

For the \textit{AI group}, the mere presence of numbers triggered heuristic reasoning that associated  mathematical representations with logical algorithmic thinking process even when they could “‘not fully understand the logic behind [NR’s] decision making.’ (A43)”~\cite{ehsan2021explainable}. 
Contradictorily, they voted the least understandable robot (NR) to be more intelligent! 
To them, “‘Math [...had] an aura of intelligence’, which ‘made the NR robot feel smarter’ (A16, A75)”~\cite{ehsan2021explainable}. 
Not only did they over-value numerical representation, but this group also viewed numbers as potentially actionable \textit{
even when} their meaning was unclear. 
Actionability refers to what one might do with the information in terms of diagnosis or predicting future behavior. 
Many highlighted that \textit{even if} they could not ‘make sense of numbers right now, [they] should be able to act on them in the future’ (A39)”~\cite{ehsan2021explainable}. 
But how \textit{actionable} were NR’s numbers in actuality? As we highlighted before, Q-values cannot indicate the “why” behind the decision. 
These numbers do not allow much actionability beyond an assessment of the quality of the actions available. 
That is, despite a desire to correct errors, having the numbers on hand would not help them determine the cause of failures that could be corrected.
Instead, they engendered over-trust and misplaced assessment of the robot’s intelligence.  

For the \textit{non-AI group}, the very inability to understand complex numbers triggered heuristic reasoning that NR must be intelligent. 
NR’s “`language of numbers', \textit{because} of its ‘cryptic incomprehensibility’, signaled intelligence (NA6, NA1)”~\cite{ehsan2021explainable}. 
Note that this line of reasoning is different from the AI group’s heuristic which posited a future actionability (despite present lack of understandability). 

Operationalizing the concept of EPs, our case study showcases how \textit{even when} there is \textit{no intention} to deceive, unsuspected effects (i.e., over-reliance on numbers) can arise. 
For either group, we did not anticipate that unlabeled, seemingly incomprehensible numbers would increase trust and assessment of the agent's intelligence. 
Moreover, we presented the Q-values in good-faith. 
What if these numbers were manipulated? 
Imagine bad-faith actions exploiting these pitfalls to manifest dark patterns.
For instance, an XAI system that explains in (manipulated) numbers (to induce trust); given the heuristic faith in numbers, it can induce over-trust and incorrect perceptions of the system.


\section{Navigation Strategies for Explainability Pitfalls}
\vspace{-5pt}
Given their nature, it is unlikely that we can completely eliminate explainability pitfalls (EPs).
Recall the uncertainty around EPs--- just because they exist does not guarantee the downstream harms will happen.
We do not yet know enough to predict when, how, and why a given AI explanation will trigger unanticipated negative downstream effects. 
While we are vulnerable to pitfalls and there is no silver bullet solution, we can 
increase our \textit{resiliency} by adopting \textit{pitfall-aware} strategies---proactive and preventative measures that help us understand where pitfalls tend to be found, how they work, and how they can be avoided. 
To expand on the pitfall metaphor, we want to probe the areas (``grounds'') of the explanation design space (where pitfalls are likely to occur) to increase our likelihood of being on sturdy ground.
We can be pitfall-aware (proactive and preventative) in our approaches at \textit{three} interconnected levels: research, design, and organizational. 

At the \textbf{research level}, we need to conduct more situated and empirically diverse human-centered research to obtain a refined understanding of the stakeholders as well as the dimensions of explanations that affect different stakeholders in XAI. 
This is because pitfalls become symptomatic---and thereby identifiable---when downstream effects (like user perceptions on AI explanations) manifest.  
For instance, the case study in Section~\ref{sec:case} revealed that different AI backgrounds in end-users can trigger the same pitfall (over-trust in numbers) but for different heuristic reasons. 
Without running the study, we could not have identified these pitfalls. 
Fortunately, the case study was a controlled lab experiment and not a real-world deployment, which limits the potential harm done by EPs.
However, the exploration in the case study revealed an important blind spot around the divergent interpretations of explanations based on one's AI background.
Building on these insights, we can do further research on a range of relevant areas.
For instance, how combinations of user characteristics (e.g., educational and professional backgrounds) impact susceptibility to EPs, how different heuristics can combine to manifest harmful biases, and how different users appropriate explanations in unexpected manners. 
Taking a pitfall-aware mindset in these explorations can generate actionable insights about how end-user reactions to AI explanations may diverge from designer intentions.

At the \textbf{design level}, we seek design strategies that are resilient to pitfalls.
One possible strategy can be to \textit{shift} our explanation design philosophy to emphasize user reflection (as opposed to acceptance) during interpretation of explanations. 
Recent human-centered XAI work has also advocated for conceptualizing ways to foster trust via \textit{reflection}~\cite{ehsan2020human}.
In terms of origins, some pitfalls are a consequence of \textit{uncritical acceptance} of explanations.
Langer et al.~\cite{langer1978mindlessness} point out that people are likely to accept explanations without conscious attention if no effortful thinking is required from them.
In Kahneman’s dual-process theory~\cite{kahneman2011thinking} terms, this means that if we do not invoke mindful and deliberative (system 2) thinking with explanations, we increase the likelihood of uncritical consumption.
To trigger mindfulness, Langer et al.~\cite{langer1978mindlessness} recommend to design for “effortful responses” or “thoughtful responding”.
To help with mindfulness, we can incorporate the lenses of {\em seamful design}~\cite{chalmers2003seamful}, which emphasize configurability, agency, appropriation, and revelation of complexity~\cite{inman2019beautiful}. 
Seamful design is the complement of the notion of "seamlessness" in computing systems (cf.~\cite{chalmers2003seamful,broll2005seamful,inman2019beautiful}) and has conceptual roots in Ubiquitous Computing~\cite{chalmers2003seamful}.

The notion of seam{\em ful}ness aligns well with XAI because (a)~AI systems are deployed in what Vertesi calls \textit{seamful} spaces~\cite{vertesi2014seamful}, and (b)~the approach can be viewed as a response to ``seamless'' black-boxed AI decisions with “zero friction” or understanding.
In terms of form and function, seams strategically reveal complexities and mechanisms of connection between different parts while concealing distracting elements. 
This notion of “strategic revealing and concealment” is central to seamful design because it connects form with function~\cite{inman2019beautiful}.
Understanding of such connections can promote reflective thinking~\cite{chalmers2003seamful}. 
\textit{Seamful explanations}, thus, \textit{strategically reveal} relevant information that augment system understanding and \textit{conceal} those that distract. 
They shed light on both the imperfections and affordances of the system, awareness of which can add \textit{useful} cognitive friction and promote effortful and reflective thinking. 
Examples of seamful explanations include interactive counterfactual explanations where we prompt the user with what-if scenarios. 
For instance, what if the AI group members were prompted to reflect using counterfactual scenarios on Q-values? 
Making the counterfactuals explicit can help the user to be aware of the variability around the system’s decision, which can help build better mental models of the system~\cite{miller2019explanation}. 
Moreover, by facilitating contrastive thinking, counterfactuals elicit engagement and deliberative thinking~\cite{byrne2019counterfactuals}, which can potentially help navigate around EPs.

At an {\bf organizational level}, we can introduce educational (training) programs (e.g., pitfall literacy programs) for both designers and end-users.
Having an ecosystem perspective is important because EPs have sociotechnical complexities; thus, we need strategies beyond the technical level.
Recent work has shown that literacy of dark patterns can promote self-reflection and mitigate harms~\cite{magnussonimproving}.
We can develop EP literacy programs that can (a) help designers become aware of how EPs might manifest and (b) empower end-users to identify being trapped in pitfalls. 
These programs can include simulation exercises using the  methodological lenses of speculative design~\cite{auger2013speculative} and reflective design~\cite{sengers2005reflective} to envision “what could go wrong”~\cite{colusso2019design} in facets of XAI systems. 
Designers can use participatory design~\cite{muller1993participatory} with end-users to gather effects of potential EPs. 
They can also utilize case studies (like ours) to think through the effects of the pitfalls. 
Insights from these programs can facilitate the navigation around pitfalls at both the design and evaluation levels of the ecosystem.

Taken together, these pitfall-aware strategies help us address EPs in a proactive and preventative manner, fostering resiliency against pitfalls.
These strategies are neither exhaustive nor normative, but take a formative step towards practically addressing the potential harmful of EPs.

\section{Concluding Reflections}
\vspace{-5pt}
Being able to appropriately classify negative impacts of AI explanations is crucial to making XAI systems safe and reliable. 
By starting the conversation about \textit{explainability pitfalls} (EPs), this paper brings conscious awareness to the (previously unarticulated) possibility of unintended negative effects of AI explanations.
By broadening the scope of harmful effects in XAI, EPs expand the dialogue that has already started around dark patterns.
The operationalization of EPs and proposed mitigation strategies provide actionable insights that can improve accountability and safety in XAI systems.

However, there are a number of open research questions, some of which we enumerate here:
\vspace{-5pt}
\begin{enumerate}
    \item How can we  develop a taxonomy of EPs to better diagnose and mitigate its negative effects?
    \item How might we use seamful explanations to account for the temporal evolution of pitfalls?
    \item How might we assess the impact of training programs to mitigate the effects of pitfalls?
\end{enumerate}
\vspace{-5pt}
We seek to learn from and with the HCI and AI communities through foundational and applied research to further develop the conceptual and practical facets of explainability pitfalls.
We believe that further understanding of where, how, and why unintended pitfalls reside in the design space of XAI can lead to improved safety and user empowerment in AI systems.

\begin{ack}
With our deepest gratitude, we thank our participants for generously investing their time in the case study. 
We are grateful to Zhiyu Lin, Sarah Wiegreffe, Amal Alabdulkarim, Becky Peng, and Kaige Xie for their feedback on the ideas. 
We are indebted to Samir Passi, Vera Liao, Larry Chan, Ethan Lee, and Michael Muller for their contributions to the case study, which informed the current work.
Special thanks to Rachel Urban for generously providing proofreading feedback. 
This project was partially supported by the National Science Foundation under Grant No. 1928586.

\end{ack}

\bibliography{main}

\begin{thebibliography}{10}

\bibitem{adadi2018peeking}
Amina Adadi and Mohammed Berrada.
\newblock Peeking inside the black-box: A survey on explainable artificial
  intelligence (xai).
\newblock {\em IEEE Access}, 6:52138--52160, 2018.

\bibitem{ehsan2019automated}
Upol Ehsan, Pradyumna Tambwekar, Larry Chan, Brent Harrison, and Mark Riedl.
\newblock Automated rationale generation: A technique for explainable ai and
  its effects on human perceptions.
\newblock In {\em Proceedings of the International Conference on Intelligence
  User Interfaces}, 03 2019.

\bibitem{guidotti2018survey}
Riccardo Guidotti, Anna Monreale, Salvatore Ruggieri, Franco Turini, Fosca
  Giannotti, and Dino Pedreschi.
\newblock A survey of methods for explaining black box models.
\newblock {\em ACM computing surveys (CSUR)}, 51(5):1--42, 2018.

\bibitem{holzinger2017we}
Andreas Holzinger, Chris Biemann, Constantinos~S Pattichis, and Douglas~B Kell.
\newblock What do we need to build explainable ai systems for the medical
  domain?
\newblock {\em arXiv preprint arXiv:1712.09923}, 2017.

\bibitem{katuwal2016machine}
Gajendra~Jung Katuwal and Robert Chen.
\newblock Machine learning model interpretability for precision medicine.
\newblock {\em arXiv preprint arXiv:1610.09045}, 2016.

\bibitem{che2016interpretable}
Zhengping Che, Sanjay Purushotham, Robinder Khemani, and Yan Liu.
\newblock Interpretable deep models for icu outcome prediction.
\newblock In {\em AMIA Annual Symposium Proceedings}, volume 2016, page 371.
  American Medical Informatics Association, 2016.

\bibitem{loftus2020artificial}
Tyler~J. Loftus, Patrick~J. Tighe, Amanda~C. Filiberto, Philip~A. Efron,
  Scott~C. Brakenridge, Alicia~M. Mohr, Parisa Rashidi, Jr~Upchurch,
  Gilbert~R., and Azra Bihorac.
\newblock {Artificial Intelligence and Surgical Decision-making}.
\newblock {\em JAMA Surgery}, 155(2):148--158, 02 2020.

\bibitem{MacKenzie2018}
Donald MacKenzie.
\newblock {Material Signals: A Historical Sociology of High-Frequency Trading}.
\newblock {\em American Journal of Sociology}, 123(6):1635--1683, 2018.

\bibitem{murawski2019mortgage}
John Murawski.
\newblock Mortgage providers look to ai to process home loans faster.
\newblock {\em Wall Street Journal}, March 2019.

\bibitem{Rudin2020}
Cynthia Rudin, Caroline Wang, and Beau Coker.
\newblock The age of secrecy and unfairness in recidivism prediction.
\newblock {\em Harvard Data Science Review}, 2(1), 3 2020.
\newblock https://hdsr.mitpress.mit.edu/pub/7z10o269.

\bibitem{Kleinberg2017}
Jon Kleinberg, Himabindu Lakkaraju, Jure Leskovec, Jens Ludwig, and Sendhil
  Mullainathan.
\newblock {Human Decisions and Machine Predictions}.
\newblock {\em The Quarterly Journal of Economics}, 133(1):237--293, 2017.

\bibitem{hao2019jail}
Karen Hao.
\newblock Ai is sending people to jail -- and getting it wrong.
\newblock {\em MIT Technology Review}, January 2019.

\bibitem{stumpf2016explanations}
Simone Stumpf, Adrian Bussone, and Dympna O’sullivan.
\newblock Explanations considered harmful? user interactions with machine
  learning systems.
\newblock In {\em ACM SIGCHI Workshop on Human-Centered Machine Learning},
  2016.

\bibitem{smith2020no}
Alison Smith-Renner, Ron Fan, Melissa Birchfield, Tongshuang Wu, Jordan
  Boyd-Graber, Daniel~S Weld, and Leah Findlater.
\newblock No explainability without accountability: An empirical study of
  explanations and feedback in interactive ml.
\newblock In {\em Proceedings of the 2020 CHI Conference on Human Factors in
  Computing Systems}, pages 1--13, 2020.

\bibitem{Kaur2020}
Harmanpreet Kaur, Harsha Nori, Samuel Jenkins, Rich Caruana, Hanna Wallach, and
  Jennifer {Wortman Vaughan}.
\newblock {Interpreting Interpretability: Understanding Data Scientists' Use of
  Interpretability Tools for Machine Learning}.
\newblock In {\em Proceedings of the 2020 CHI Conference on Human Factors in
  Computing Systems}, CHI '20, pages 1--14, New York, NY, USA, 2020.
  Association for Computing Machinery.

\bibitem{ghai2021explainable}
Bhavya Ghai, Q~Vera Liao, Yunfeng Zhang, Rachel Bellamy, and Klaus Mueller.
\newblock Explainable active learning (xal) toward ai explanations as
  interfaces for machine teachers.
\newblock {\em Proceedings of the ACM on Human-Computer Interaction},
  4(CSCW3):1--28, 2021.

\bibitem{Eiband2019}
Malin Eiband, Daniel Buschek, Alexander Kremer, and Heinrich Hussmann.
\newblock {The Impact of Placebic Explanations on Trust in Intelligent
  Systems}.
\newblock In {\em Extended Abstracts of the 2019 CHI Conference on Human
  Factors in Computing Systems}, CHI EA '19, pages 1--6, New York, NY, USA,
  2019. Association for Computing Machinery.

\bibitem{brignull2015dark}
Harry Brignull, Marc Miquel, Jeremy Rosenberg, and James Offer.
\newblock Dark patterns-user interfaces designed to trick people, 2015.

\bibitem{chromik2019dark}
Michael Chromik, Malin Eiband, Sarah~Theres V{\"o}lkel, and Daniel Buschek.
\newblock Dark patterns of explainability, transparency, and user control for
  intelligent systems.
\newblock In {\em IUI workshops}, volume 2327, 2019.

\bibitem{gray2018dark}
Colin~M Gray, Yubo Kou, Bryan Battles, Joseph Hoggatt, and Austin~L Toombs.
\newblock The dark (patterns) side of ux design.
\newblock In {\em Proceedings of the 2018 CHI Conference on Human Factors in
  Computing Systems}, pages 1--14, 2018.

\bibitem{zagal2013dark}
Jos{\'e}~P Zagal, Staffan Bj{\"o}rk, and Chris Lewis.
\newblock Dark patterns in the design of games.
\newblock In {\em Foundations of Digital Games 2013}, 2013.

\bibitem{capurro2021dark}
Daniel Capurro and Eduardo Velloso.
\newblock Dark patterns, electronic medical records, and the opioid epidemic.
\newblock {\em arXiv preprint arXiv:2105.08870}, 2021.

\bibitem{ehsan2021explainable}
Upol Ehsan, Samir Passi, Q~Vera Liao, Larry Chan, I~Lee, Michael Muller, Mark~O
  Riedl, et~al.
\newblock The who in explainable ai: How ai background shapes perceptions of ai
  explanations.
\newblock {\em arXiv preprint arXiv:2107.13509}, 2021.

\bibitem{watkins1992q}
Christopher~JCH Watkins and Peter Dayan.
\newblock Q-learning.
\newblock {\em Machine learning}, 8(3-4):279--292, 1992.

\bibitem{kahneman2011thinking}
Daniel Kahneman.
\newblock {\em Thinking, fast and slow}.
\newblock Macmillan, 2011.

\bibitem{wason1974dual}
Peter~C Wason and J~St~BT Evans.
\newblock Dual processes in reasoning?
\newblock {\em Cognition}, 3(2):141--154, 1974.

\bibitem{petty1986elaboration}
Richard~E Petty and John~T Cacioppo.
\newblock The elaboration likelihood model of persuasion.
\newblock In {\em Communication and persuasion}, pages 1--24. Springer, 1986.

\bibitem{ehsan2020human}
Upol Ehsan and Mark~O Riedl.
\newblock Human-centered explainable ai: Towards a reflective sociotechnical
  approach.
\newblock In {\em International Conference on Human-Computer Interaction},
  pages 449--466. Springer, 2020.

\bibitem{langer1978mindlessness}
Ellen~J Langer, Arthur Blank, and Benzion Chanowitz.
\newblock The mindlessness of ostensibly thoughtful action: The role of"
  placebic" information in interpersonal interaction.
\newblock {\em Journal of personality and social psychology}, 36(6):635, 1978.

\bibitem{chalmers2003seamful}
Matthew Chalmers and Ian MacColl.
\newblock Seamful and seamless design in ubiquitous computing.
\newblock In {\em Workshop at the crossroads: The interaction of HCI and
  systems issues in UbiComp}, volume~8, 2003.

\bibitem{inman2019beautiful}
Sarah Inman and David Ribes.
\newblock " beautiful seams" strategic revelations and concealments.
\newblock In {\em Proceedings of the 2019 CHI Conference on Human Factors in
  Computing Systems}, pages 1--14, 2019.

\bibitem{broll2005seamful}
Gregor Broll and Steve Benford.
\newblock Seamful design for location-based mobile games.
\newblock In {\em International Conference on Entertainment Computing}, pages
  155--166. Springer, 2005.

\bibitem{vertesi2014seamful}
Janet Vertesi.
\newblock Seamful spaces: Heterogeneous infrastructures in interaction.
\newblock {\em Science, Technology, \&amp; Human Values}, 39(2):264--284, 2014.

\bibitem{miller2019explanation}
Tim Miller.
\newblock Explanation in artificial intelligence: Insights from the social
  sciences.
\newblock {\em Artificial Intelligence}, 267:1--38, 2019.

\bibitem{byrne2019counterfactuals}
Ruth~MJ Byrne.
\newblock Counterfactuals in explainable artificial intelligence (xai):
  Evidence from human reasoning.
\newblock In {\em IJCAI}, pages 6276--6282, 2019.

\bibitem{magnussonimproving}
Jonathan Magnusson.
\newblock Improving dark pattern literacy of end users.

\bibitem{auger2013speculative}
James Auger.
\newblock Speculative design: crafting the speculation.
\newblock {\em Digital Creativity}, 24(1):11--35, 2013.

\bibitem{sengers2005reflective}
Phoebe Sengers, Kirsten Boehner, Shay David, and Joseph'Jofish' Kaye.
\newblock Reflective design.
\newblock In {\em Proceedings of the 4th decennial conference on Critical
  computing: between sense and sensibility}, pages 49--58, 2005.

\bibitem{colusso2019design}
Lucas Colusso, Cynthia~L Bennett, Pari Gabriel, and Daniela~K Rosner.
\newblock Design and diversity? speculations on what could go wrong.
\newblock In {\em Proceedings of the 2019 on Designing Interactive Systems
  Conference}, pages 1405--1413, 2019.

\bibitem{muller1993participatory}
Michael~J Muller and Sarah Kuhn.
\newblock Participatory design.
\newblock {\em Communications of the ACM}, 36(6):24--28, 1993.

\end{thebibliography}
\bibliographystyle{unsrt}

\end{document}